\documentclass[amssymb,aps,byrevtex,superscriptaddress]{revtex4}
\usepackage{epsfig,graphics,graphicx}% Include figure files
\usepackage{color}

\begin{document}

\title{Two-dimensional dissipative rogue waves due to time-delayed feedback in cavity nonlinear optics}
\author{Mustapha Tlidi}
\affiliation{Facult{\'e} des Sciences,Optique Nonlin\'eaire Th\'erorique, Universit{\'e} libre de B ruxelles (U.L.B.), C.P.
231, Campus Plaine, B-1050 Bruxelles, Belgium}
\author{Krassimir Panajotov}
\affiliation{Department of Applied Physics and Photonics (IR-TONA),  Vrije Universiteit Brussels,  Pleinlaan 2,
B-1050 Brussels,  Belgium}
\affiliation{Institute of Solid State Physics,  72 Tzarigradsko Chaussee Blvd.,  1784 Sofia,  Bulgaria}

\begin{abstract}
We demonstrate a way to generate  a two-dimensional rogue waves  in  two types of broad area nonlinear optical systems subject to time-delayed feedback: in the generic Lugiato-Lefever model and   in model of a broad-area surface-emitting laser with saturable absorber.  The delayed feedback is found to induce a spontaneous formation of rogue waves. In the absence of delayed  feedback, spatial pulses are stationary. The rogue waves are exited and controlled by  the delay feedback. We characterize their formation by computing the probability distribution of the pulse height. The  long-tailed statistical contribution which is often considered as a signature of the presence of rogue waves appears for sufficiently strong feedback. The generality of our analysis suggests that the feedback induced instability leading to the spontaneous formation of two-dimensional rogue waves is an universal phenomenon.
\end{abstract}

\maketitle
{\textbf{Optical rogue waves may be generated by fibers and appear with a very large amplitude comparing to other surrounding pulses. They consist of optical pulses that appear and disappear suddenly. The long tail probability distribution is the fundamental characteristics accounting for the generation of rogue waves. The theory of rogue waves was mainly developed in the framework of focusing nonlinear Schroedinger equation. However, the nonlinear  Schroedinger equation does not admit two-dimensional solutions due to collapse dynamics. In this contribution, we consider a two-dimensional rogue waves in cavity nonlinear optics described by the well known Lugiato-Lefever model equation. In particular, we discuss the effect of a delay feedback on the generation of  two-dimensional rogue waves in the transverse plane of the cavity. In the absence of the delay effect, the  two-dimensional pulses are stationary in time. When increasing the strength of the delay feedback, the pulse undergoes instabilities that lead to the formation of rogue waves. }}

\section{Introduction} 
Spatial and/or temporal confinement of light leading to the formation of localized structures (LS's) are drawing considerable attention both from fundamental as well as from applied point of views. These stationary solutions occur in a dissipative environment and belong to the class of dissipative structures.  Their  existence, due to the occurrence of a Turing (or modulational) instability, has been abundantly discussed and is by now fairly well understood (see recent overviews on this issue \cite{TTK_2007,Nail_lect,Leblond-Mihalache,Clerc,Tlidi-PTRA,Lugiatobook,Tlidi_Clerc_sringer}).
Along another line of research, theoretical and experimental studies have shown the possibility to generate rogue waves (RW) in fiber optics.  Rogue waves are rare giant pulses  or extreme events.  They are also called dissipative rogue waves and have been generated in passively mode-locked lasers \cite{Akhmediev_2011}. The long tail probability distribution is the fundamental characteristics accounting for the generation of rogue waves. The modulational instability is an important mecanism for the creation of ocean waves as shown by Peregrine \cite{Peregrine}. In addition,  Peregrine solitons are considered as a prototype of rogue waves formation. Experimental confirmation of Peregrine solitons has been demonstrated in optical fiber \cite{Mussot_08,Kibler} and in water wave tank \cite{Chabchoub,chabchoub2} systems. Small amplitude pulses may grow to large amplitudes if their frequencies fall in the band of unstable mode with a positive gain. Nonlinear interaction between unstable frequencies may lead to a very complicated wave dynamics. Analytical study of the nonlinear interaction between two frequencies solutions of the nonlinear Schroedinger equation in the form of the collision of Akhmediev breathers has been reported in \cite{Nail}.

The formation of RW in optics has been the subject of intense research since the pioneering work by Solli et al. \cite{Solli}. Since then, the number of systems in which rogues waves are identified to appear has become important and can be witnessed by recent review papers \cite{Akhmediev_13,Onorato,Dudley_14,Akhmediev_2016}. Recently, several studies have indicated that spatially extended systems  exhibit rogue waves in the transverse section  of the  light \cite{Odent,Arecchi,Birkholz,Montica,Lushnikov,Marshal,Leonetti,Pierangeli}. Rogue wave may be generated in anisotropic inhomogeneous nonlinear medium \cite{Zhong}.

The purpose of this paper is to report that the appearance of two-dimensional rogue waves in dissipative systems:  driven optical cavities subjected to optical injection and broad-area surface-emitting lasers with saturable absorber. We propose a mechanism to generate  two-dimensioal rogue waves formation by time delayed feedback scheme. This mechanism has been recently applied successfully in  one-dimensional systems \cite{Tlidi_Intropy,Akhmediev_2016}. In the first part of the present paper, we focus on two-dimensional Lugiato-Lefever model. We show that depending on the strength of the delay feedback, localized structures become unstable and rogue waves are formed. The rogue wave formation can occur in the regime where the transmitted intensity as a function of the input intensity is monostable.  We provide a statistical analysis showing a non-Gaussian profile of the probability distribution with a long tail and pulse intensity height well beyond two times the significant wave height (SWH). The SWH is  defined as the mean height of the highest third of waves. In the second part, we perform the same analysis by using a model describing broad-area surface-emitting laser with saturable absorber.  The generality of our analysis suggests that the instability leading to the spontaneous formation of rogue waves is an universal phenomenon.  It is worth to mention an important theoretical work on rogue waves by Akhmediev et al. \cite{Akmhediev_OL_09} in the framework on Schroedinger nonlinear equation. However, we choose two systems that cannot be described by the Nonlinear Schroedinger equation. This is  because the nonlinear Schroedinger equation does not admit two-dimensional solutions due to collapse dynamics.

In the case of small area semiconductor laser where the diffraction is neglected, there exist a narrow parameter regions where the laser intensity exhibits high intensity pulses in the time domain \cite{Tredicci}. When the delay feedback is taken into account temporal rogue waves have been also generated \cite{Reinoso}. Rogue waves as extreme events have been reported in one dimensional systems such as all fiber cavities \cite{Conforti} and whispering gallery mode resonators \cite{Chembo14}. More recently, an analysis of rogue waves supported by experimental data
in a  semiconductor microcavity laser with intracavity
saturable absorber has been reported \cite{Selmi}. In all these works, the generation of rogue waves have been studied in strictly one dimensional setting.

In this paper, after an introduction, we characterize in section II, the formation of rogue waves generated in two-dimensional Lugiato-Lefever Model with a delay feedback. In Section III, we consider a model describing broad-area surface-emitting laser with a saturable absorber with a time delay feedback and we show the occurrence of two-dimensional rogue waves in this system. We conclude in section VI.

\section{Lugiato-Lefever model} 
We propose for the first time a mechanism of the formation of two-dimensional  spatial rogue waves based on the  time delayed feedback control scheme. To illustrate this mechanism  we consider a nonlinear passive cavity subjected to time-delayed feedback. This system is modeled by the well known Lugiato-Lefever (LL) equation that describes a ring cavity filled with a Kerr media and driven coherently by an injected signal \cite{LL}.  This model equation is valid under the
following approximations:~the cavity possesses a high Fresnel number, \textit{i.e.} it is a large-aspect-ratio system and
we assume that the cavity is much shorter than the diffraction and the nonlinearity
spatial scales; and for the sake of simplicity, we assume a single longitudinal mode
operation. We implement in the LL model an optical time-delayed feedback as a single round-trip delay term \cite{Tlidi_prl09,Panajotov_epjd10}. More precisely, we adopt the  Rosanov \cite{rosanov75} and in the Lang and Kobayashi \cite{LK} approach to model the time-delay feedback. The LL model with optical feedback reads:
\begin{eqnarray}
\frac{dE}{dt} = ia\nabla^{2}_{\perp} E  - (\alpha+i\theta)E + i|E|^2E + E_i +\\ \nonumber
\eta e^{i\phi}\left[E(t-\tau)-E(t)\right].\label{eq:dEdt}
\end{eqnarray}
Here $E=E(x,y,t)$ is the normalized mean-field cavity electric field and $\theta$ is the frequency detuning parameter.  $\alpha$ is the cavity losses, and $E_i$ is the input field amplitude assumed to be real, positive
and independent of the transverse coordinates. Diffraction is modeled by the Laplace  operator $\nabla^{2}_{\perp}= \partial^2 /\partial x^{2}+\partial^2 /\partial y^{2}$ acting on the transverse plane $(x,y)$. The diffraction coefficient is $a$. For simplicity we normalized the losses to $\alpha=1$, like in classical Lugiato-Lefever equation. In this case, the homogeneous steady states are $I_i = I_S\left[ 1+(\theta-I_S)^2 \right] I_S$, with $I_i=E_i^2$ and $I_S=|E_S|^2$. Depending on choice of $\theta$ the curve $I_S(I_i)$ is either monostable ($\theta<\sqrt{3}$) or bistable ($\theta>\sqrt{3}$) \cite{LL}. The delay feedback is modelled by an external cavity operating in a self-imaging configuration \cite{Tlidi_prl09,Panajotov_epjd10}. This allows us to compensate diffraction in the external cavity. This feedback  is characterized by the time-delay $\tau$, feedback strength $\eta$ and phase $\phi$. In the absence of delay feedback, the LL model has been derived for other systems such as all fiber cavity \cite{Healterman} and  whispering gallery mode resonators \cite{Chembo}. In these two systems the diffraction term modeled by the Laplace operator is replaced by chromatic dispersion effect modeled by a second derivative with respect to the retarded time in the reference frame moving with the group velocity of light.  The LL model has also been derived for a cavity with left-handed materials \cite{Kockaert.2006,Lendert07} where diffraction coefficient is negative $a<0$.

\begin{figure}[t!]
\centerline{\includegraphics[width=9cm]{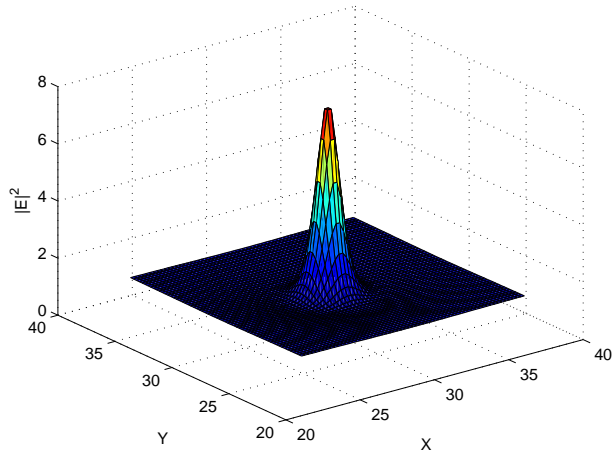}}
\caption{Localized structure in the 2D Lugiato-Lefever model without optical feedback. Parameters are $\theta=1.7$ and $E_i=1.2$.}
\label{fig:CS}
\end{figure}
It is well known that in the absence of delay feedback, i.e. $\eta=0$, the Lugiato-Lefever equation admits stationary localized structures in one and two-dimensional settings \cite{Scroggie} similar to those reported in \cite{TML94}. Experimental evidence of spatial LS in Kerr medium has been realized in \cite{Odent.2011,Odent-RC}.

We fix all the parameters in Lugiato-Lefever model and only vary the strength of the delay feedback. In the absence of delay feedback, i.e. $\eta=0$, a single or multipeak stationary localized structures are formed (see Fig.~\ref{fig:CS}). The occurrence of stationary localized structures does not require a bistable homogeneous response curve. The  prerequisite condition for their formation is the coexistence between a single homogeneous flat solutions and the spatially periodic pattern. This coexistence occurs in the monostable regime \cite{Scroggie}, i.e., for $\theta<\sqrt{3}$. A weakly nonlinear analysis in one \cite{LL} and in two transverse dimensions \cite{TML96} have shown that this coexistence occurs when $\theta> 41/30$. In this regime, the relative stability analysis has shown that the only stable periodic pattern in two-dimensions are hexagons, other symmetries are unstable \cite{TML96}. The interaction of well-separated localized structures has been also discussed \cite{VTZ_2012}

When increasing the value of $\eta\tau$ above $1$, such localized structures exhibit a regular drift with a constant velocity (not shown) that has been reported in \cite{Krassimir_2016}. The delay feedback allows for the motion of stationary localized structure when the product $\eta\tau$ reaches the value of $+1$ for  $\phi=\pi$ \cite{Tlidi_prl09,Panajotov_epjd10,Pimenov_pra13}. Optical delay feedback may also be at the origin of a drift
bifurcation leading to the motion of localized strctures in the Lugiato-Lefever equation \cite{Krassimir_2016}.

\begin{figure}[h!]
\centerline{\includegraphics[width=8cm]{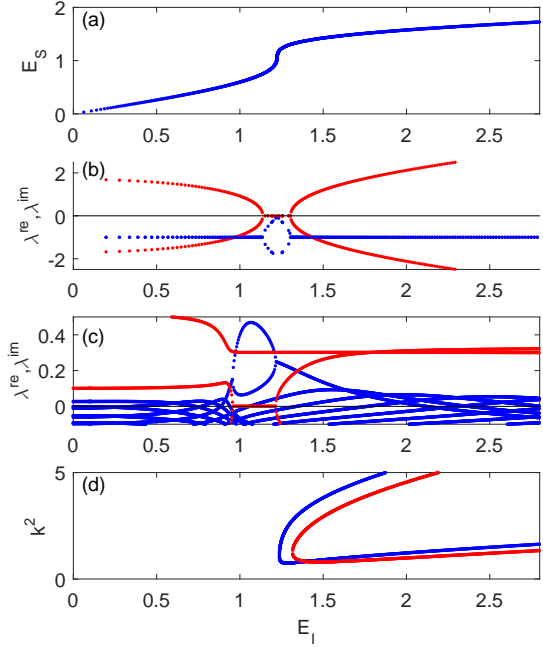}}
\caption{(color online) Lugiato-Lefever model homogeneous steady-state solution $E_S(E_i)$ (a) and its stability for $\eta=0.0$ (b) and for $\eta=0.7$ and $\phi=3\pi/4$ (c). In (b) and (c) the real and imaginary parts of the eigenvalues are shown by blue and red color, respectively. (d) $k^2$ at the onset of Turing instability: blue (red) are for the case of $\eta=0$ and $\eta=0.7$, $\phi=3\pi/4$, respectively. LL parameters are the same as in Fig.~\ref{fig:CS}.}
\label{fig:HSS}
\end{figure}

%%%%%%%%%%%%%%%%%%%%%%%%%%%%%%%%%%
The linear stability of the homogeneous steady states (HSS) is analyzed by considering small fluctuations around the steady-state that are modulated with transverse wavevector $k_{\bot}$. The optical feedback impacts the stability of the homogeneous solution by both its magnitude and phase. This is illustrated in Fig.~\ref{fig:HSS} for the case of monostable HSS. Without optical feedback the HSS is stable with respect to spatially-homogeneous perturbations ($k_{\bot}=0$) [see Fig.~\ref{fig:HSS} (b)] and Turing (modulationally) unstable above $E_i\sim1.24$. The optical feedback drastically changes the stability creating a plead of Hopf bifurcations and thus making the HSS unstable in the whole range of $E_i$ as depicted in Fig.~\ref{fig:HSS} (c). It also modifies the region of Turing instabilities [see Fig.~\ref{fig:HSS} (d)].
%
%%%%%%%%%%%%%%%%%%%%%%%%%%%%%%
\begin{figure}[t!]
%\centerline{\includegraphics[width=9cm]{Roque_Wave_2D_snapshot.eps}}
\centerline{\includegraphics[width=9cm]{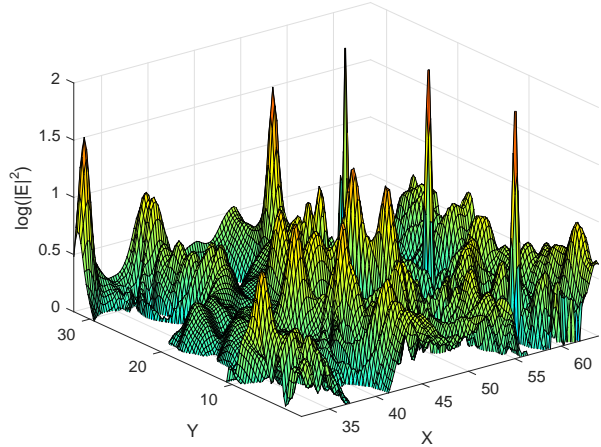}}
\caption{A snapshot of the optical intensity in logarithmic scale for the 2D Lugiato-Lefever model with an extreme event captured. The parameters are the same as in Fig.~\ref{fig:CS} and the optical feedback parameters are $\eta=0.7$, $\tau=10$, and $\phi=3\pi/4$.}
\label{fig:EE}
\end{figure}

\begin{figure}[t!]
\centerline{\includegraphics[width=9cm]{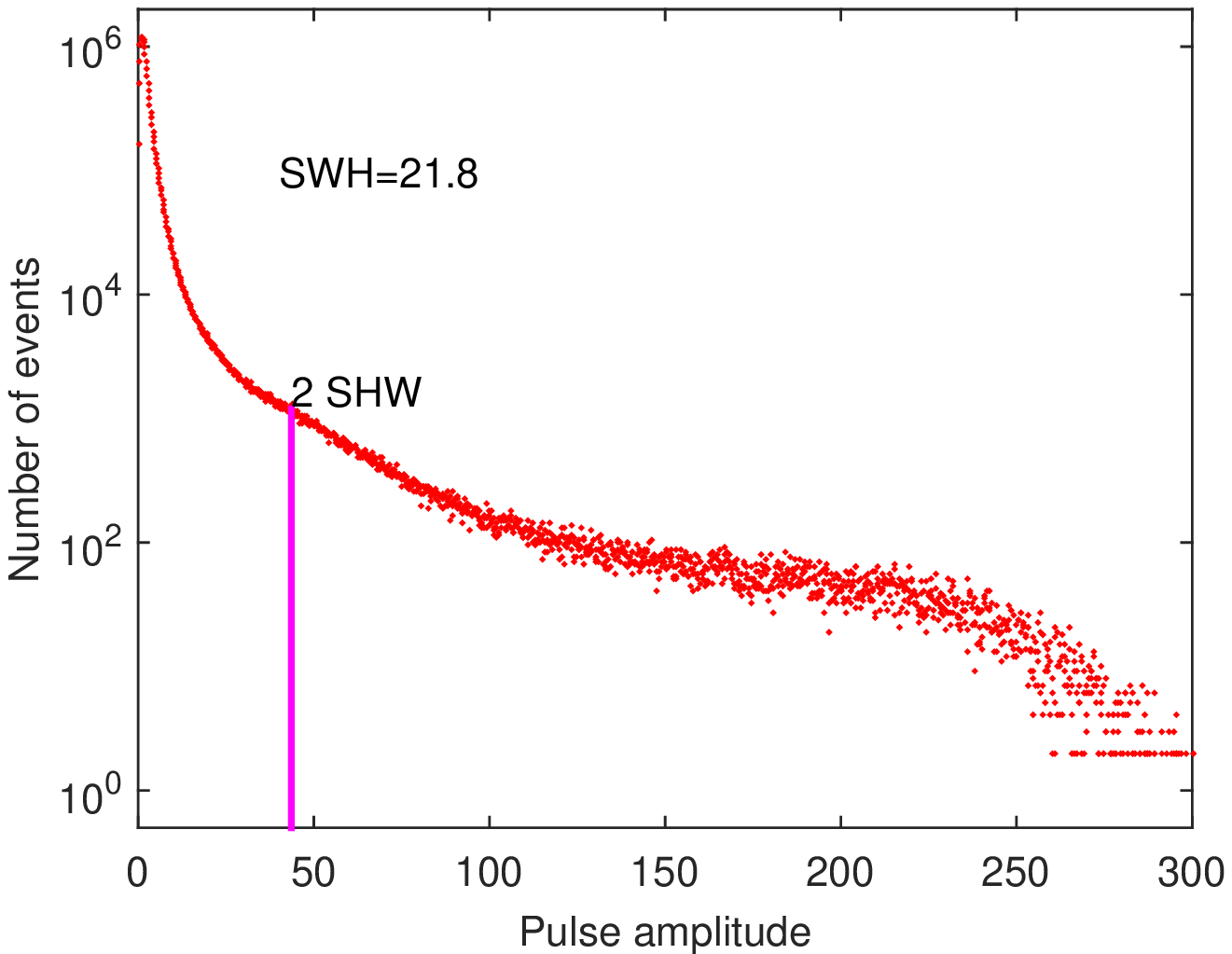}}
\caption{The number of events as a function of the intensity of the pulses in semi-logarithmic scale.  Parameters are the same as in Fig.~\ref{fig:EE}. The SWH denotes the significant wave height.  The dashed line indicates 2 x SWH.}
\label{fig:distr}
\end{figure}
When further increasing the value of $\eta$, spatial-temporal chaos appears which further develops so that very high amplitude pulses (extreme events) appear (see the snapshot of the optical intensity shown in Fig.~\ref{fig:EE} for $\eta=0.7$, $\tau=10$, and $\phi=3\pi/4$). A statistical analysis shows that the height of such extreme events is more than twice the significant wave height (SWH) - see Fig. \ref{fig:distr}. This figure shows a non Gaussian statistics of the wave intensity, with a long tail of the probability distribution typical for rogue waves formation.

\section{Broad-area surface-emitting laser with a saturable absorber} 
Another problem which produces spatial rogue waves is the broad-area surface-emitting laser with a saturable absorber. Recently, spatiotemporal chaos and extreme events have been demonstrated experimentally in an extended microcavity laser in 1D configuration \cite{Selmi}. Here, we consider the control of two-dimensional rogue waves by time-delayed optical feedback. We consider the mean field model describing the space-time evolution of broad area vertical cavity surface emitting laser (VCSEL) with saturable absorption \cite{Bache_apb05} and modify it by adding a delay optical feedback from a distant mirror in a self-imaging configuration, i.e. light diffraction in the external cavity is compensated  \cite{Panajotov_ol14}:
\begin{eqnarray}
\frac{dE}{dt} &=& \left[\left(1-i\alpha\right)N + \left(1-i\beta\right)n - 1 + i\nabla^{2}_{\perp}\right]E \\ \nonumber &+& \eta e^{i\phi}E(t-\tau), \label{eq:dEdt} \\
\frac{dN}{dt} &=& b_1 \left[\mu - N\left(1 + \left|E\right| ^{2}\right)\right], \label{eq:dNdt}\\
\frac{dn}{dt} &=& -b_2 \left[\gamma + n\left(1 + s\left|E\right| ^{2}\right)\right]. \label{eq:dndt}
\end{eqnarray}
Here $E$ is the slowly varying  mean electric field envelope. $N$ ($n$) is related to the carrier density, $\alpha$ ($\beta$) is the linewidth enhancement factor and $b_1$  ($b_2$) is the ratio of photon lifetime to the carrier lifetime in the active layer (saturable absorber) (normalization is the same as in \cite{Bache_apb05}). $\mu$ is the normalized injection current in the active material,  $\gamma$ measures absorption in the passive material and $s=a_2b_1/(a_1b_2)$ is the saturation parameter with $a_{1(2)}$ the differential gain of the active (absorptive) material. The diffraction of intracavity light $E$ is described by the Laplace operator $\nabla^{2}_{\perp}$ acting on the transverse plane $(x,y)$ and carrier diffusion and bimolecular recombination are neglected. Time and space are scaled to the photon lifetime $\tau_p$ and diffraction length, respectively. The feedback is characterized by the time-delay $\tau$, the feedback strength $\eta$ and phase $\phi$.

We consider the same laser parameters as in \cite{Panajotov_ol14}: $\alpha=2$, $\beta=0$, $b1=0.04$, $b2=0.02$, $\gamma=0.5$, $s=10$, $\mu=1.42$ and optical feedback with a time-delay of $\tau=100$ and phase $\phi=0$. The current is chosen such that the laser without optical feedback resides in a bistable region between the zero homogeneous solution ($E=0$, $N=\mu$, $n=-\gamma$) and the lasing solution ($E=\sqrt{I}e^{iwt}$, $N=\mu/(1+I)$, $n=-\gamma/(1+sI)$). For this choice of parameters the upper branch exhibits a subcritical Turing type of bifurcation allowing for the formation of LSs, which experiences a period-doubling bifurcation to spatially localized chaos \cite{Panajotov_ol14}. The time-delayed feedback parameters are: $\tau=100$, $\eta=0.75$ and $\phi=0$. We integrate numerically Eqs. (1)-(\ref{eq:dndt}) by using the standard split-step method with periodic boundary conditions.

\begin{figure}[t!]
\centerline{\includegraphics[width=9cm]{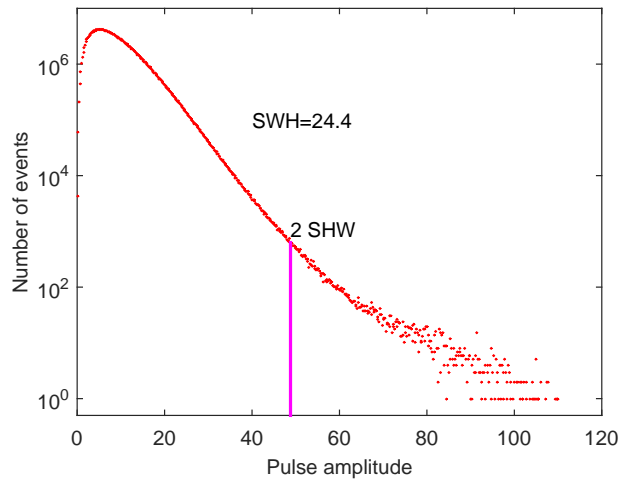}}
\caption{Statistical distribution of pulse height of spatial-temporal chaos induced by optical feedback in the model of a broad-area surface-emitting laser with a saturable absorber. The number of events as a function of the amplitude of the pulses in semi-logarithmic scale.  The SWH denotes the significant wave height.  The dashed line indicates 2 x SWH.}
\label{fig:SA_distr}
\end{figure}
Statistical analysis of pulse height distribution of spatial-temporal chaos in the model of a broad-area surface-emitting laser with a saturable absorber is presented in Fig. \ref{fig:SA_distr}. The  long-tailed statistical contribution serves as a signature of the presence of rogue waves: rogue waves with pulse heights more than twice the SWH appear in the system.
%color red

\section{Conclusion}
 We demonstrate a way to generate rogue waves by time-delayed feedback in two generic nonlinear systems: a broad area nonlinear optical resonator subject to optical injection and a broad-area surface-emitting laser with a saturable absorber.  While in the absence of delayed  feedback the spatial pulses are stationary, for sufficiently strong feedback spontaneous formation of rogue waves is observed.  These rogue waves are clearly exited and controlled by  the feedback. The generality of our analysis suggests that the feedback induced instability leading to the spontaneous formation of rogue waves is an universal phenomenon.

\begin{acknowledgments}
M.T. received support from the  Fonds National de la Recherche Scientifique (Belgium). We aknowledge the support of   the Interuniversity
Attraction Poles program Photonics@be of the Belgian Science Policy Office (BelSPO), under Grant IAP\,7-35.  K.P. is grateful to the Methusalem foundation for financial support.
\end{acknowledgments}

\end{document}